\documentclass[conference]{IEEEtran}
\IEEEoverridecommandlockouts
\usepackage{cite}
\usepackage{amsmath,amssymb,amsfonts}
\usepackage{algorithmic}
\usepackage{graphicx}
\usepackage{textcomp}
\usepackage{subcaption} 
\graphicspath{{Breast_Cancer_Images/}}
\usepackage{xcolor}
\def\BibTeX{{\rm B\kern-.05em{\sc i\kern-.025em b}\kern-.08em
		T\kern-.1667em\lower.7ex\hbox{E}\kern-.125emX}}

\begin{document}
\title{A Deep Analysis of Transfer Learning Based Breast Cancer Detection Using Histopathology Images }
\author{\IEEEauthorblockN{Md Ishtyaq Mahmud}
	\IEEEauthorblockA{\textit{College of Science and Engineering} \\
		\textit{Central Michigan University}\\
		Mount Pleasant, MI 48858, USA \\
		mahmu4m@cmich.edu}
	\and
	\IEEEauthorblockN{Muntasir Mamun}
	\IEEEauthorblockA{\textit{Department of Computer Science} \\
		\textit{University of South Dakota}\\
		South Dakota, USA \\
		muntasir.mamun@coyotes.usd.edu}
	\and
	\IEEEauthorblockN{Ahmed Abdelgawad}
	\IEEEauthorblockA{\textit{College of Science and Engineering} \\
		\textit{Central Michigan University}\\
		Mount Pleasant, MI 48858, USA \\
		abdel1a@cmich.edu}}
\maketitle

\begin{abstract} 
	Breast cancer is one of the most common and dangerous cancers in women, while it can also afflict men. Breast cancer treatment and detection are greatly aided by the use of histopathological images since they contain sufficient phenotypic data. Deep Neural Network (DNN) is commonly employed to improve accuracy and breast cancer detection. In our research, we have analyzed pre-trained deep transfer learning models such as  ResNet50, ResNet101, VGG16, and VGG19 for detecting breast cancer using 2453 histopathology images dataset. Images in the dataset were separated into two categories: those with invasive ductal carcinoma (IDC) and those without IDC. After analyzing the transfer learning model,  we found that ResNet50 outperformed other models, achieving accuracy rates of 90.2\%, Area under Curve(AUC) rates of 90.0\%, recall rates of 94.7\%, and a marginal loss of 3.5.

\end{abstract}

\IEEEpeerreviewmaketitle


 \section{Introduction}
 Cancer is the result of uncontrolled irregular cell proliferation. The most frequent disease to affect women, but it can also affect men, is breast cancer, which has a high rate of diagnosis. Breast cancer is highly preventable and curable when caught early, which lowers the mortality rate \cite{9091012}. Breast cancer is one of the deadliest cancers considering the statistical data of  American Cancer Societies . The American Cancer Society estimates that in 2022 more than 287,850 new patient will be diagnosed and the death rate  will be 43,780 for breast cancer in the USA \cite{Cancer}. The World Health Organization (WHO) report on cancer in 2020 is presented in this article\cite{mdpi,9965732,10019058}. Figure 1 shows  estimated statistical data of some types of cancer in 2022.
 
 \begin{figure}[htbp]
 	\centering
 	\includegraphics[width=.45\textwidth]{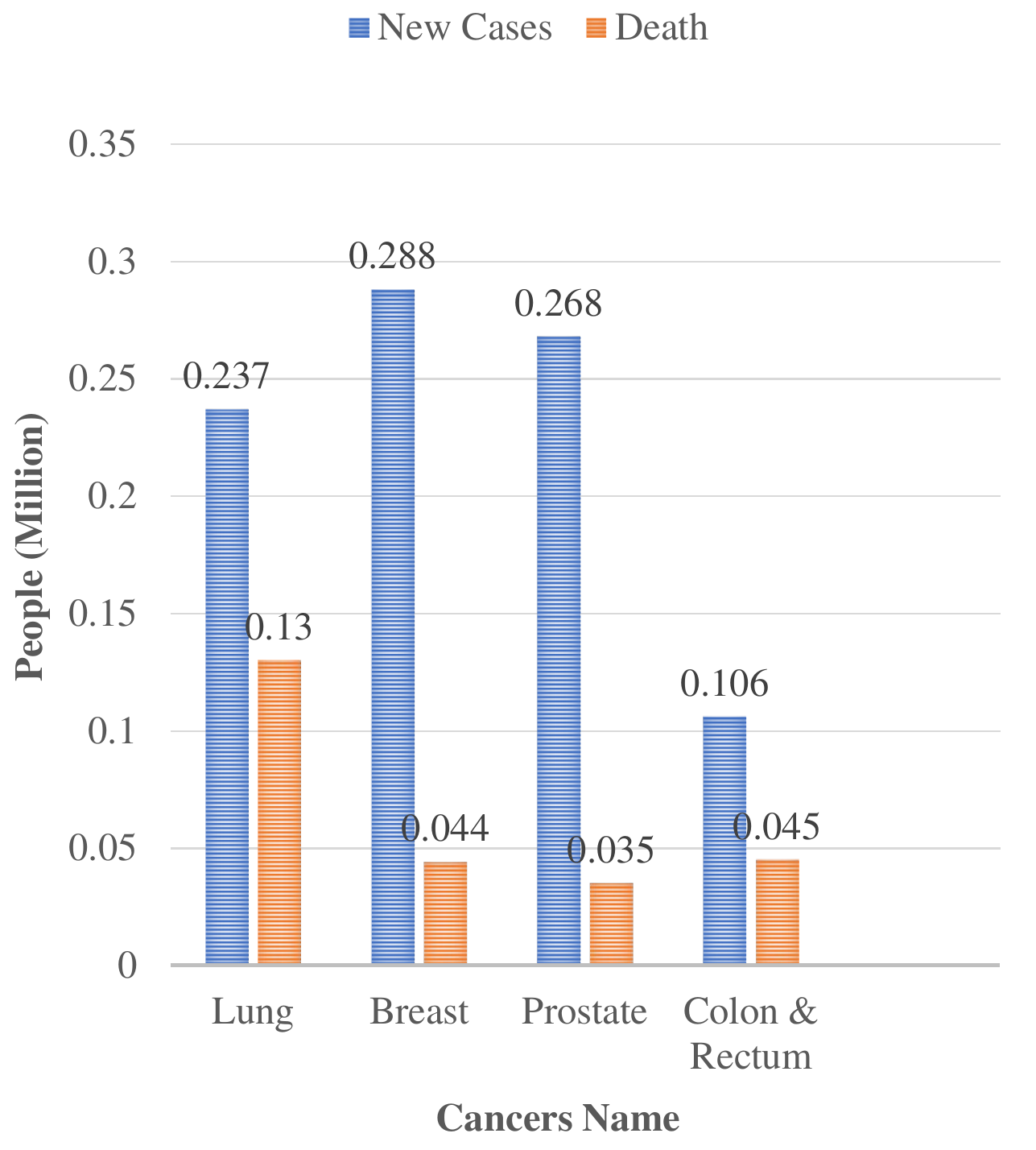}
 	\caption{Cancer in 2022 (Estimate new cases against death)}
 	\label{Fig.1.}
 \end{figure}

 In-situ carcinoma, invasive carcinoma, normal, and benign are the four different forms of breast tissue. Benign tissue does not provide a health risk because it can produce minimal changes to the breast's structure. In-situ cancer stays in the breast duct lobule system and does not spread to other organs. In-situ cancer can be treated if it is discovered in time. However, invasive carcinoma is a cancerous tumor that has a propensity to invade more organs. According to \cite{nassif2022breast} there are several methods used for detecting breast cancer including X-ray mammography, breast temperature measurement, Portion Emission Tomography (PET), Computed Tomography (CT), ultrasound (US), and Magnetic Resonance Imaging (MRI). Also, for the diagnosed, two approach can be used: genomics and histopathology\cite{9965730}.  Histopathology images, which are microscopic pictures of breast tissue, are very helpful in the early stages of cancer treatment. A new area of study called radio-genomics focuses on multi-scale correlations between gene expression data and medical imaging \cite{sutanto2015benchmark}.
 
 To recent advances in machine learning, it is now possible to detect serious health issues and provide beneficial results. Deep learning is frequently utilized in the healthcare industry for early disease identification and diagnosis. Also, Deep learning in particular has made great strides in the field of image interpretation by making it simpler to identify, classify, and quantify patterns in images of the body \cite{9817303, MHfit}. In order to analyze deep learning models for identifying and diagnosis  breast cancer, infrared or histopathology images are typically used \cite{nemade2022review, Lcd}.
  
 
 Following is how the remaining portion of the article is organized: Section II lists the prior works. The methodology was shown in Section III, where we provided specifics on dataset collection, dataset processing, validation procedure, and pre-trained transfer learning strategy. Performance of the transfer learning model is distributed in section IV and summarizes the conclusion and future works are provided in section V.

\section{Related Works}

This section provides a detailed description of the related works, and Table I lists the findings of earlier studies on the detection of breast cancer.

Chaves et al. \cite{chaves2020evaluation} introduced transfer learning methods (AlexNet, GoogLeNet, ResNet-18, VGG-16, and VGG-19) for detecting breast cancer. To getting promising results they used 440 infrared thermography images of 88 patients. They got 77.5 \% accuracy with both VGG-16, and VGG-19. 

Allugunti et al. \cite{allugunti2022breast} proposed Computer-aided Diagnosis (CAD) methods for analyzing and diagnosis breast cancer patient. The researcher also applied different classifier of machine learning such as, CNN, the Support Vector Machine (SVM), and Random Forest (RF). They employed mammography pictures, which helped them achieve a greater classification accuracy rate. Comparing to the authors results, CNN (99.67\%) performed better than others.

 Goncalves et al. \cite{gonccalves2022cnn} proposed two distinct forms of bio-inspired technologies to enhance CNN's design. They applied VGG-16, ResNet-50 and DensNet-201 to improve F1 score. Their applied three modern CNN networks received F1 scores above 0.90. Two classes of patients' infrared photographs from the DMR-IR database were used.
 
  Farooq et al. \cite{farooq2020infrared} proposed inception-v3 model for detecting and analyzing breast cancer using thermal medical images. They used DMR-IR datset where data was collected from 287 patients with different age. They obtained 80\% overall accuracy, F1 score 76.89\% and 83.33\% sensitivity with their deep neural network model.
  
   Cabioglu et al. \cite{cabiouglu2020computer} introduced deep transfer learning model for diagnosis and treatment of breast cancer using thermal images. AlexNet worked better for their research. Using transfer learning methods with CNN, they were able to achieve accuracy, precision, and recall of 94.3\%, 94.7\%, and 93.3\%, respectively.
   
   Roslider \cite{8875661} et al. introduced pre-trained transfer learning model for breast cancer detection. They used various types of transfer learning model such as ResNet-101, DenseNet-201 MobileNet-V2 and
   ShuffleNet-V2. Among them, DenseNet-201 outperformed the others and had 100\% accuracy.
   
    Amrane et al. \cite{amrane2018breast} proposed two ddifferent ML classifier such as   Naive Bayes (NB) classifier and knearest neighbor (KNN) for detecting breast cancer. The ML classifier KNN outperforms the NB classifier (96.19\%) in accuracy while achieving a lower error rate (97.51\%). The study employed data from Wisconsin Breast Cancer databases, which included 683 patients overall, 444 benign patients, and 239 malignant patients.
    
    Michael et al. \cite{michael2022optimized} introduced five different ML classifier such as  k-NN, SVM, RF, XGBoost, and LightGBM  for detecting breast cancer. Accuracy, precision, recall, and F1-score for the LightGBM classifier were 99.86\%, 100.00\%, 99.60\%, and 99.80\%, respectively, better than those of the other four classifiers. In the dataset, there were 912 ultrasound images total, 600 of which were benign and 312 of which were cancer.
    
    In our research work for breast cancer detection we use four various types of  pre-trained transfer learning model  namely, Resnet50,, Resnet101, VGG 16 and VGG19. Some of the resercher used infrared images for their research, though we used 2453 histopathology breast images for analyzing our model and got promising results for ResNet50. We use feature extraction for processing the images. We also compared the results of Accuracy, loss, recall and AUC considering our transfer learning model. ResNet50 performed much better than other models, with accuracy rates of 90.2\%, area under curve (AUC) rates of 90.0\%, recall rates of 94.7\%, and a minor loss of 3.5\%.
    

\begin{table}[h!]
	\centering
	\caption{Breast cancer prediction model's performance}
	\begin{tabular}{ | p{1cm} | p{1.25cm} | p{2cm}| p{2.75cm} |}
		\hline
		Reference & Dataset Collection & Models & Performance \\
		\hline
		
		\cite{chaves2020evaluation}& DMR-IR & AlexNet, GoogleNet, ResNet-18, VGG-16, and VGG-19 &  VGG16 and VGG19 : accuracy 77.5 \%  \\ 
		\hline
		
		\cite{allugunti2022breast}  & kaggle & CNN, Support Vector Machines (SVM), and Random Forest & Accuracy:  CNN 99.67\%,
		SVM 89.84\%, 
		RF 90.55\% \\
		\hline
		\cite{gonccalves2022cnn} & DMR-IR  & VGG16, ResNet-50 and DenseNet-201 & VGG-16: F1 score 0.92, ResNet-50: F1 score 0.90 \\
		\hline
		\cite{farooq2020infrared} & DMR-IR & Inception-V3  & 76.89\% F1 score,
		 80\% accuracy, and 83.33\%  sensitivity\\
		 \hline
		 \cite{cabiouglu2020computer} & DMR-IR & AlexNet & 94.3\% accuracy, precision 94.7\% and a recall 93.3\% \\
		 \hline 
		 
		 \cite{8875661} & DMR-IR & ResNet-101, DenseNet-201 MobileNet-V2 and
		 ShuffleNet-V2 & 100\% accuracy with DenseNet-201 \\
		 \hline
		 \cite{amrane2018breast} & Wisconsin Breast Cancer& Naive Bayes (NB)
		 classifier and knearest neighbor (KNN)& KNN: accuracy is 97.51\%,
		  NB: accuracy is 96.19\% \\
		 \hline 
		 \cite{michael2022optimized} & Own data (Local Hospital) &KNN, SVM, RF, XGBoost, and LightGBM & LightGBM: 99.86\% accuracy, 100\% precision , 99.60\% recall  and  99.80\%  F1 score\\
		 \hline 
		 Our Work & kaggle (Breast Histopathology Images)& ResNet50, ResNet101, VGG19 and VGG19 & ResNet50: accuracy is 90.20\%, AUC is 90\%, recall is 94.7\% and loss is 3.5 \\
		 \hline 
	\end{tabular}

\end{table}

\section{Methodology}
The methodology starts with gathering histopathology images from sources that are accessible before moving on to the pre-processing stage. Then, using the standard hold-out validation approach, pre-trained transfer learning models such RestNet-50, ResNet-101, VGG-16, and VGG-19 are trained, tested, and validated on the histopathology images dateset. By analyzing and processing the results, the most suitable transfer learning model for the detection of breast cancer can be identified. Figure 2 demonstrated the overall procedure.
\begin{figure}[htbp]
	\centering
	\includegraphics[width=.50\textwidth]{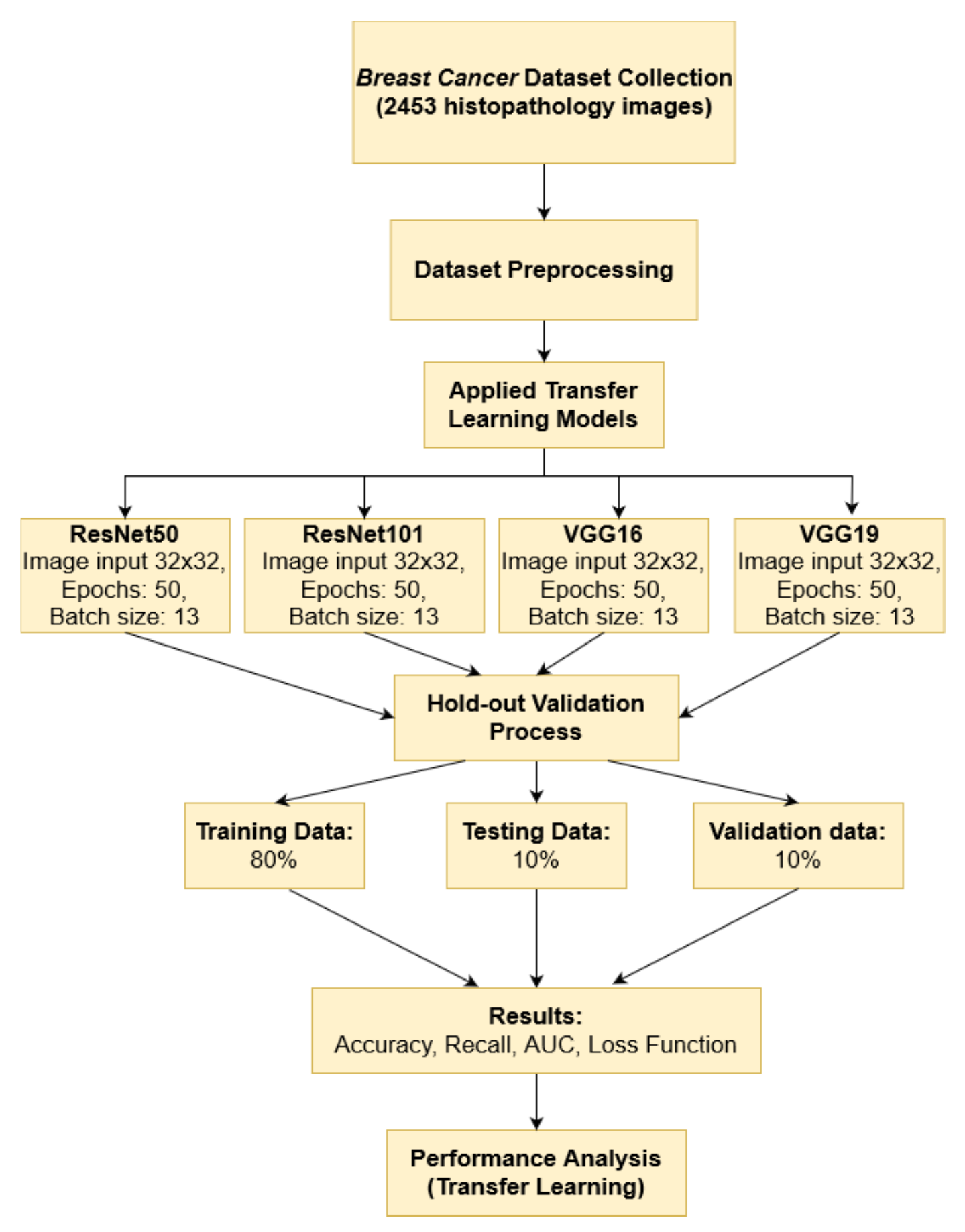}
	\caption{The Overview of the Study}
	\label{Fig.1.}
\end{figure}
 
 \begin{figure}[htbp]
	\centering
	\includegraphics[width=.45\textwidth]{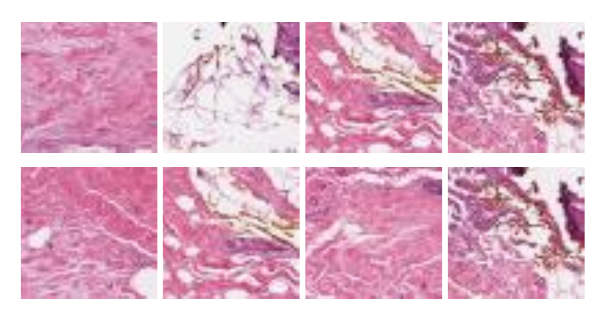}
	\caption{ Non-Cancerous Breast Histopathology Images}
	\label{Fig.1.}
\end{figure}

 \begin{figure}[htbp]
	\centering
	\includegraphics[width=.45\textwidth]{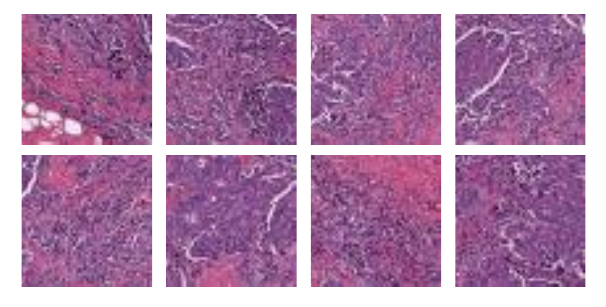}
	\caption{Cancerous Breast Histopathology Images}
	\label{Fig.1.}
\end{figure}

 \subsection{Dateset Collection}
The data set for breast cancer (Breast Histopathology Images) in this study is taken from a publicly accessible internet source called Kaggle \cite{janowczyk2016deep}. 162 whole mount slide photos of breast cancer samples scanned at 40x made up the initial dataset,  accordingly to the dataset source \cite{cruz2014automatic}. Then images were extracted and divided into two categories, Invasive Ductal Carcinoma (IDC) positive and IDC negative. The most frequent subtype of breast cancer is IDC.  We Used 2453 histopathology images for our research. Figure 3 showed the non-cancerous breast histopathology images and Figure 4 showed cancerous breast histopathology images.

 \subsection{Dataset Pre-processing}
Pre-processing is a stage that mostly involves processing the data to make it usable for training purposes. Histopathology image pre-processing is based on feature extraction using image reading, resizing, de-noising, image segmentation, and  smoothing edges. Delineating the precise IDC areas inside of a whole mount slide is a typical pre-processing procedure for the BC histopathology image dataset.  \cite{cruz2014automatic}. A pre-processing stage is vital for obtaining better accuracy after applying a transfer learning model for image classification or detection.
 \subsection{Validation Process}
 The validity process is significant to the overall study since it is important to choose the right validity procedure for a particular dataset. We used hold-out validation techniques because they produced the best results \cite{dwork2015reusable}. We divided our data into three categories: training (80\%), testing (10\%), and validation (10\%) using the hold-out process.

\subsection {Pre-Trained Transfer Learning Approaches}		
 In image classification or recognition, transfer learning is one of the most widely used techniques in machine learning. In this study we applied four types of pre-trained transfer learning model such as ResNet50, ResNet-101, VGG 16 and VGG 19 to detecting breast cancer. The following is a precise description of those transfer learning models: 
 \subsubsection{ResNet50 and ResNet101}
 ResNet is a shortened version of residual networks\cite{he2016deep} are designed with the primary goal of utilizing shortcut connections to skip entire blocks of convolutional layers. In order to address the performance deterioration and increased error that occurs as additional layers are added, a shortcut link is provided to each pair of three-by-three filters when creating a residual block. A number of advancements in the field of image recognition and classification have been accomplished over time using deep convolutional neural networks.Increasing image recognition accuracy and completing challenging tasks have become popular, however it might be challenging to train deeper neural networks due to deterioration and disappearing ingredient concerns\cite{reddy2019transfer}. Both of these issues are attempted to be resolved by residual learning. ResNet50 provides access to a 50 weighted layer residual network. ResNet101 and ResNet152 are two further variations of residual network work besides ResNet50 \cite{rezende2017malicious}.

\begin{figure}[htbp]
	\centering
	\includegraphics[width=.45\textwidth]{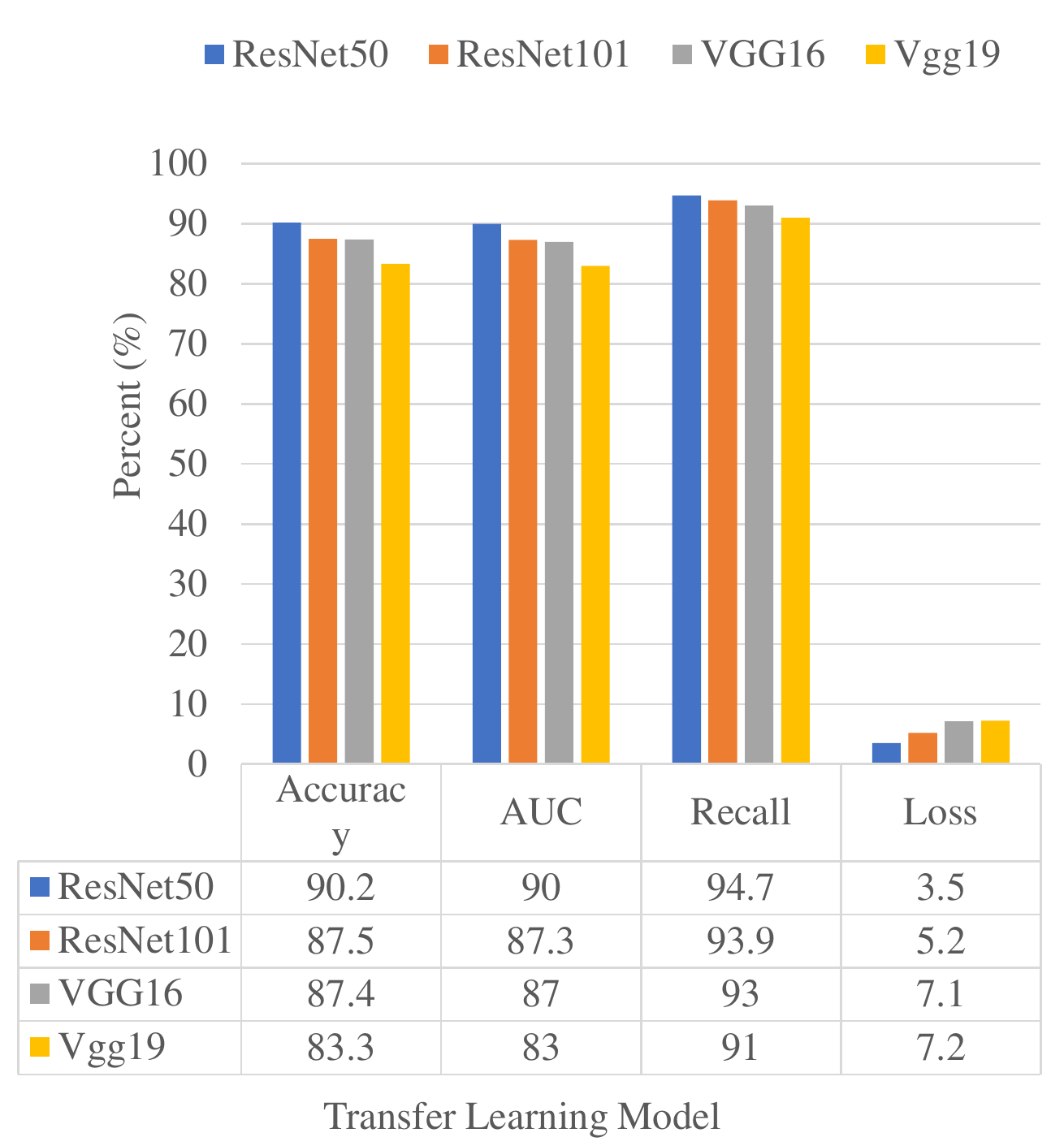}
	\caption{Results of different measures for different pre-trained transfer learning models for detecting breast cancer disease.}
	\label{FIG: performance}
\end{figure}

\subsubsection{VGG16 and VGG19}
The ImageNet Large-Scale Visual Recognition Challenge (ILSVRC) competition received a proposal for the VGG network from the Oxford Visual Geometry Group \cite{simonyan2014very}. VGG ranks among the top-performing models in the ImageNet classification challenge, which is composed of over 14 million photos divided into 1000 categories \cite{ullah2020detection} \cite{simonyan2014very}. The feature extractor VGG pre-trained on the Imagenet dataset is capable of extrapolating images quite successfully and can be utilized for a variety of applications, including images classification, and detection \cite{guan2017breast}. The VGG has different architectures, VGG16 and VGG19: the former has 16 weight layers, and the latter has 19 layers. VGG16 contains 16 layers, with the first 13 being convolutional, the next 3 being fully connected, the next 4being max-pooling layers that lower volume size and softmax activation function, and the final layer being fully connected layer. Another variation of the VGG model is the VGG19 architecture, which consists of 16 convolutional neural networks, 3 fully connected layers, 5 MaxPool layers, and 1 SoftMax layer \cite{theckedath2020detecting}.

\begin{figure*}[htbp]
	\centering
	\includegraphics[width=1\textwidth]{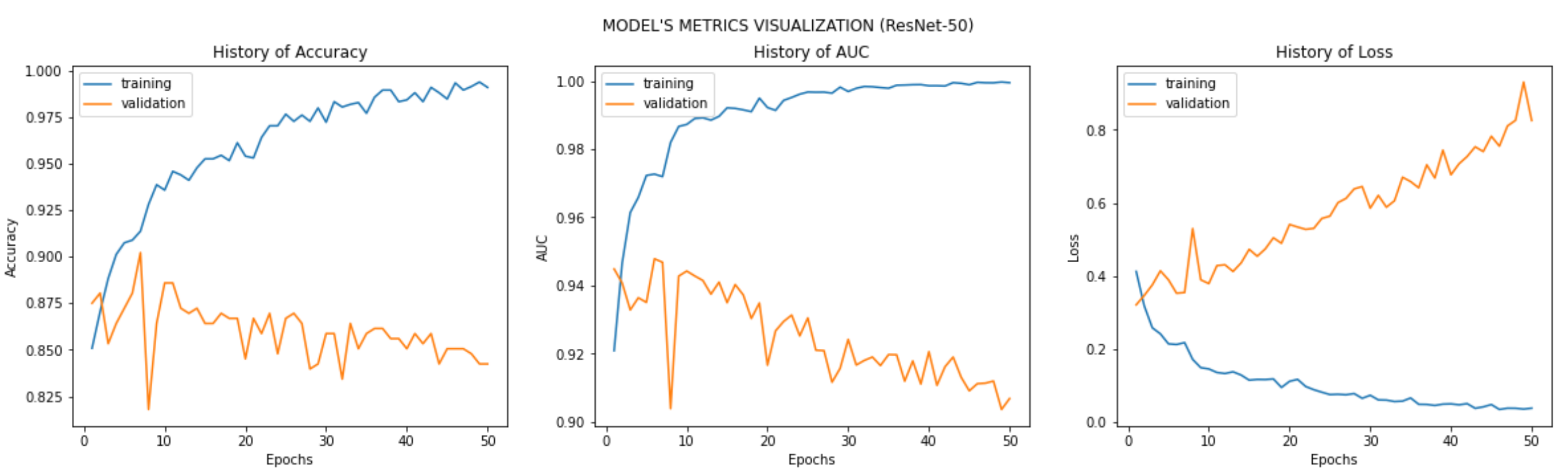}
	\caption{ResNet50 Model's Metrics Visualization}
	\label{FIG: ResNet50}
\end{figure*}

\begin{figure*}[htbp]
	\centering
	\includegraphics[width=1\textwidth]{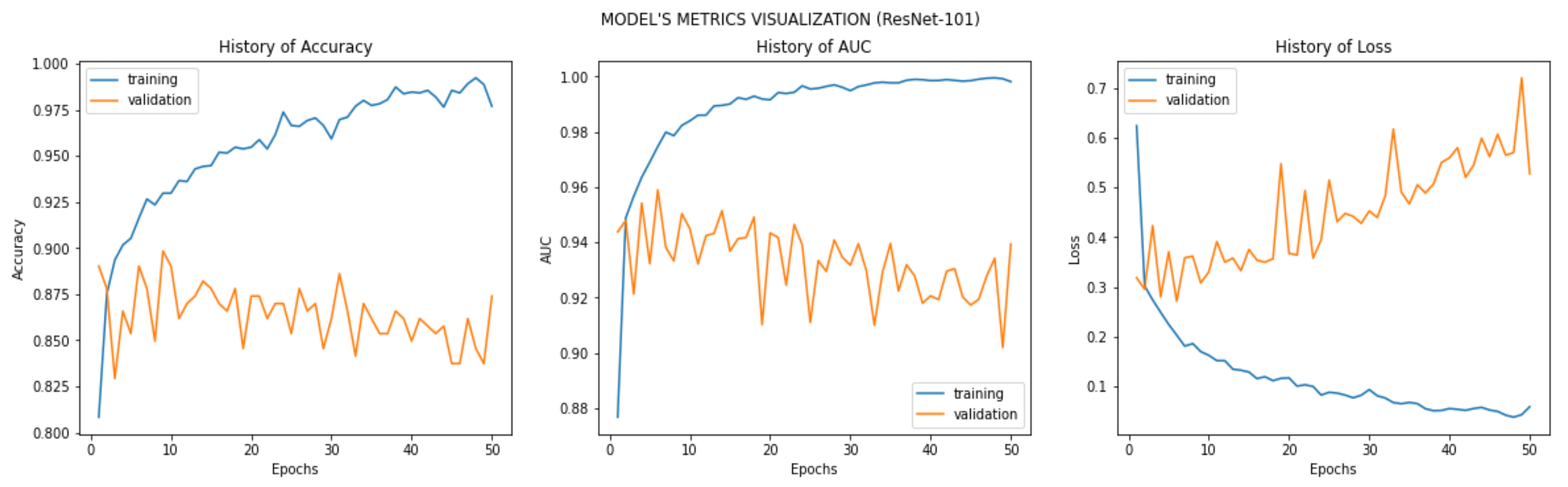}
	\caption{ResNet101 Model's Metrics Visualization}
	\label{FIG: ResNet101}
\end{figure*}

\begin{figure*}[htbp]
	\centering
	\includegraphics[width=1\textwidth]{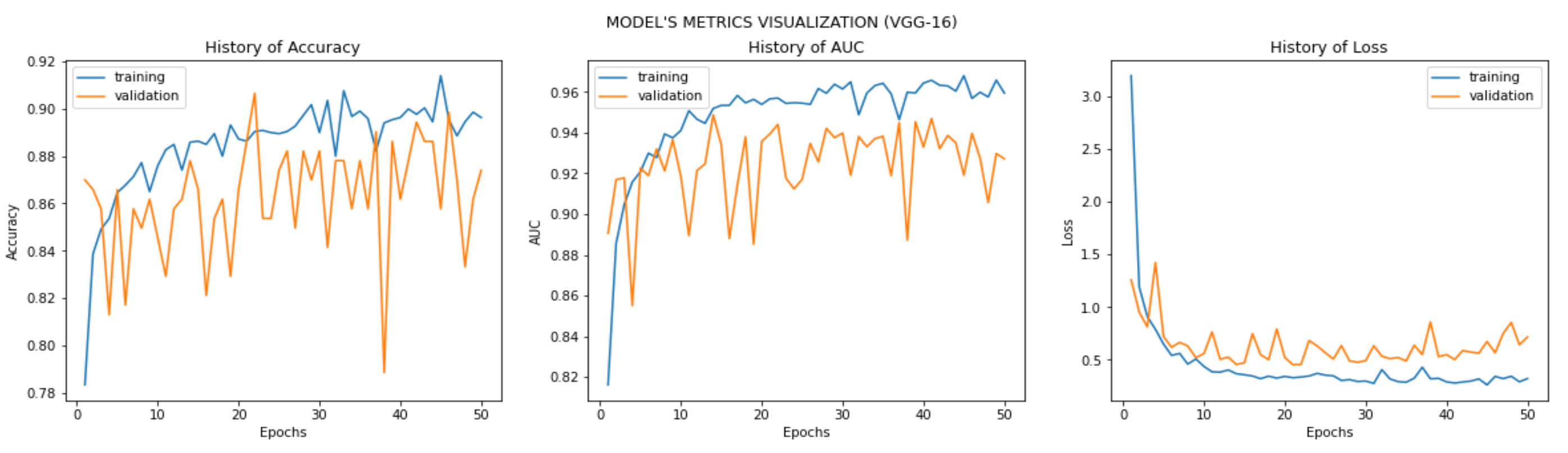}
	\caption{VGG16 Model's Metrics Visualization}
	\label{FIG: VGG16}
\end{figure*}
\begin{figure*}[htbp]
	\centering
	\includegraphics[width=1\textwidth]{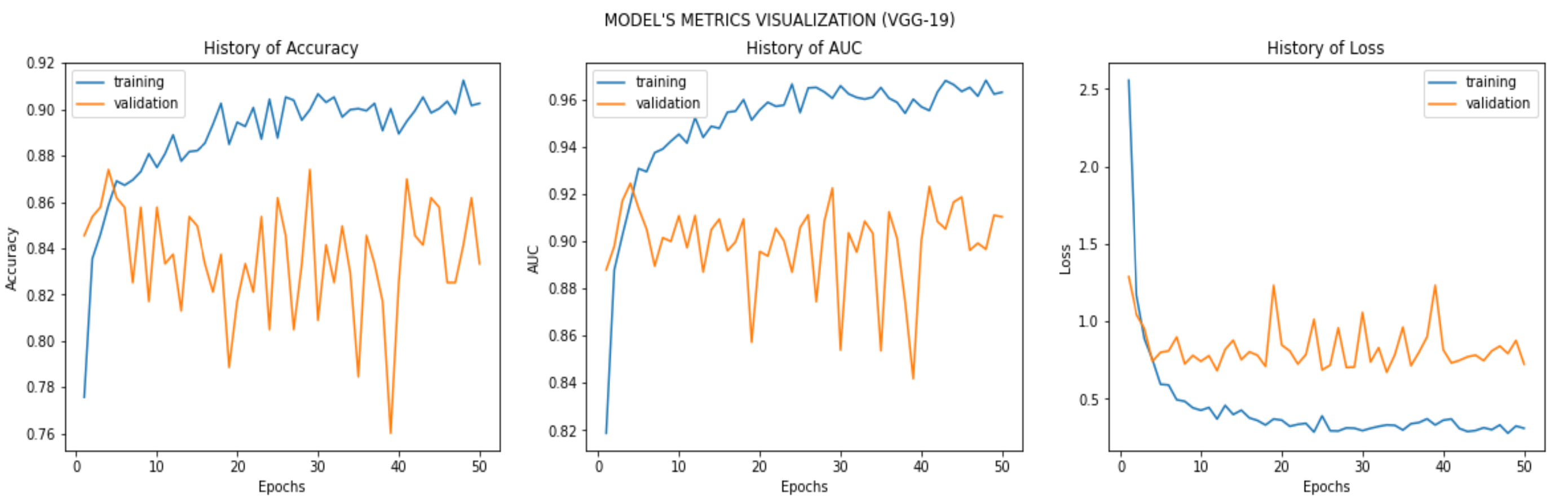}
	\caption{VGG19 Model's Metrics Visualization}
	\label{FIG: VGG19}
\end{figure*}

\section{Results and Discussions}
The performance of ResNet50, ResNet101, VGG16 and VGG19 models was evaluated to see which one performed best at detecting breast cancer based on histopathology images. The findings of Accuracy, Area Under Curve (AUC), Recall, and Loss for the performance observation of the models are presented in Fig. \ref{FIG: performance}.

We attempted to take into account four performance matrices, such as accuracy, AUC, recall, and loss, in order to choose the best pre-trained transfer learning model out of the four. We can determine which model is optimal for the detection of breast cancer by considering their performances. Accuracy by itself cannot demonstrate a competent measurement technique for assessing a model's performance.  In addition to determining the performance of the model and classifying a model, the AUC value, recall and loss function are vital. The AUC evaluates how effectively the model distinguishes between positive and negative categories. 
If the AUC values is higher, it meaning that the model performance is better.

Fig. \ref{FIG: ResNet50} , Fig. \ref{FIG: ResNet101} , Fig. \ref{FIG: VGG16}, Fig. \ref{FIG: VGG19} shows the model matrices visualizations of ResNet50, ResNet101, VGG16, and VGG19. ResNet50 outperformed other transfer learning models that had already been trained when all performance matrices were taken into consideration.As shown in Fig. \ref{FIG: ResNet50}, ResNet50 not only outperformed the competition in terms of accuracy but also achieved exceptional AUC, recall, and loss which are 90.0 \%, 94.7\% and 3.5\%. The loss function is a significant additional indicator for evaluating the model's performance  The loss function provides information on the variance from training and validation values. Our loss function values typically have high values if the validation value is too far off from the training value. Lower loss values represent improved model performance. In terms of loss matrices, ResNet50 has a loss of about 3.5\%, which is lower than that of other models like ResNet101, VGG16, and VGG19, which are 5.2, 7.1, and 7.2, respectively. With the exception of the loss function, which is 5.2 \% for ResNet101 and 7.1 \% for VGG16, we have achieved almost similar performance for accuracy, AUC, and recall for ResNet101 and VGG16. The overall performance for identifying breast cancer using VGG19 is the weakest out of four pre-trained transfer learning models, with 83.3\% accuracy, 83.0\% AUC, 91.0\% recall and 7.2 loss.

\section{Conclusion and Future work}
A growing number of women are being diagnosed with breast cancer, and it is one of the most deadly diseases. Breast cancer can be cured if caught early, and in many circumstances, it can even be prevented. Using infrared or histopathological pictures, a machine learning algorithm can be applied to determine the presence of breast cancer. In this research we use breast histopathology images for detecting breast cancer. We obtained the dataset from kaggle, which contains 2453 breast histopathology images. The images were processed based on feature extraction. For the purpose of detecting breast cancer, we have presented pre-trained deep transfer learning models including ResNet50, ResNet101, VGG16, and VGG19. In the future, we can employ sophisticated pre-trained transer learning models to detect breast cancer with more accuracy. In order to perform the model, we can also take a large amount of dataset into consideration.

\bibliographystyle{IEEEtran}
\bibliography{Bibliography}

\end{document}